\documentstyle[aps,twocolumn]{revtex}
\begin{document}
\twocolumn[\hsize\textwidth\columnwidth\hsize\csname @twocolumnfalse\endcsname
\title{Robustness of a Local Fermi Liquid against Ferromagnetism and 
Phase Separation II}
\author{K.S. Bedell, J.R. Engelbrecht, and K.B. Blagoev$^*$}
\address{Department of Physics, Boston College, Chestnut Hill, MA 02167}
\date{\today}
\maketitle
\begin{abstract}
This is a response to a recent paper by van Dongen,
Uhrig, and E. M\"uller-Hartmann.
We clarify the issues, which the above authors 
apparently did not fully appreciate, 
by referring to
the microscopic foundations of 
Landau's Fermi Liquid Theory, its general application to
the theory of metals and the theory of local Fermi Liquids.
\end{abstract}
]
In a recent paper\cite{van Dongen et al.1998} van Dongen, Uhrig,
and M\"uller-Hartmann 
express their concerns with
our local Fermi Liquid Theory 
\cite{Engelbrecht and Bedell1995,Blagoev et al.1998a}
and its implications for
local theories of paramagnetic and ferromagnetic metals. 
Their concerns with our work
may be 
based on pathologies in the infinite-$d$
many-body theories on which the above authors have written many 
papers in the past few years.
Several incorrect statements made by van Dongen, 
Uhrig, and M\"{u}ller-Hartmann in \cite{van Dongen et al.1998}
can be traced back to a confusion by the authors, regarding
the microscopic foundations of
Fermi Liquid Theory.

Before answering their specific objections to our work, we would
like to make a 
historic
comment. 
The concept of a local Fermi Liquid Theory, pre-dates the recently developed
infinite-dimensional theories by nearly four decades.
Textbook examples are
the electron-phonon problem\cite{Migdal1958,Schrieffer1964} and the
electron-paramagnon problem\cite{Berk and Schrieffer1966,Berk1966}.
local Fermi Liquid Theory is exact in infinite dimensions, but it
can and has been  
applied to three dimensions. 
In fact our ``ongoing research''\cite{Blagoev et al.1998a,Blagoev et al.1998b} 
is just
a continuation of this line of thought in the case of ferromagnetic
metals. 
 
There are three specific points, raised by van Dongen, 
Uhrig, and M\"{u}ller-Hartmann. Here we
discuss these points in some detail, although they have been in
textbooks for more than thirty years\cite{Abrikosov et al.1963}.

Their first point of criticism is that 
``Engelbrecht and Bedell
{\it assume} without justification that their local Fermi Liquid is
{\it isotropic}''. 
We are somewhat perplexed since in Ref.~\cite{Engelbrecht and Bedell1995}
we formally proved that 
for a local (i.e. momentum-independent) self-energy,
Landau's $f$-function is independent of the momenta
on the Fermi Surface,
\begin{equation}
f_{\sigma\sigma^\prime}({\bf p},{\bf p^\prime})=A+B\sigma\cdot\sigma^\prime.
\end{equation}
The constants $A$ and $B$ are indeed invariant under rotations of 
${\bf p}$ and ${\bf p^\prime}$.
Next van Dongen et al. explain that in infinite $d$:
``For arbitrary $\delta >0$ one can always find pairs 
$({\bf p},{\bf p'})$ with ${\bf p}$ on the Fermi surface 
($\varepsilon({\bf p})=\varepsilon_{{\tiny\mbox{F}}}$) and ${\bf p'}$ close
to ${\bf p}$ (i.e., $|{\bf p'}-{\bf p}|/|{\bf p}|<\delta$), while
$|\varepsilon ({\bf p'})-\varepsilon_{{\tiny\mbox{F}}}|$ is large (comparable
to the bandwidth)''. 
This means that the energy is not a 
continuous function of the momentum. While in principle it
cannot be ruled out on physical grounds it is rather
pathological and not the expected situation for three dimensional metals. 

Their second major objection concerns the local nature of the
vertex function. There is a major confusion on behalf of van Dongen, 
Uhrig, and M\"{u}ller-Hartmann. 
They apparently mixed the full 4-point vertex function 
$\Gamma_{p,p^\prime}(q)$
with the irreducible vertex function for zero-momentum transfer, i.e.
$\Gamma^{IR}_{p,p^\prime}(q=0)$ (we use momentum-energy
notations, i.e. $q=(\bf{q},\omega)$). It is well known that,
\cite{Abrikosov et al.1963}
\begin{equation}
2i\frac{\delta\Sigma(p)}{\delta G(p^\prime)}=\Gamma^{IR}_{p,p^\prime}(q=0),
\end{equation}
with $\Sigma(p)$ the self-energy and $G(p)$ the single particle
Green's function. A sufficient condition for the existence of a $\Phi$
functional (so called $\Phi$-derivable theory) is that 
$\Gamma^{IR}_{p,p^\prime}(q=0)$ is a symmetric function of $p$ and $p^\prime$. 
This is discussed in more detail 
in chapter four in the 1963 Dover edition of the Abrikosov, Gorkov, and
Dzyaloshinskii book\cite{Abrikosov et al.1963}. It then follows that
if the self-energy does not depend on $\bf{p}$, then 
$\Gamma^{IR}_{p,p^\prime}(q=0)$ does not depend on $\bf{p}$ or 
$\bf{p}^\prime$. With the authors apparently 
confusing $\Gamma^{IR}_{pp^\prime}(q=0)$ and
$\Gamma_{pp^\prime}(q)$ their comments regarding the lack of momentum dependence
in our local Fermi liquid are understandable. The full $\Gamma_{pp^\prime}(q)$
as well as our dynamic susceptibility clearly depend on $q=(\bf{q},\omega)$
(for further details see Ref.~\cite{Abrikosov et al.1963}).

In their next comment van Dongen, 
Uhrig, and M\"{u}ller-Hartmann question Eq.(6) in our paper. This equation
relates the Fermi Liquid parameters $F_0^s$ and $F_0^a$. It is based
on two exact, fundamental relations in Fermi liquid 
theory\cite{Landau1957,Baym and Pethick1991} namely the relation between
the Fermi Liquid parameters, $F_l^{s,a}$, and the scattering amplitudes 
\begin{equation}
A_l^{s,a}=\frac{F_l^{s,a}}{1+F_l^{s,a}/(2l+1)},
\end{equation}
and the Landau (forward scattering) sum rule
\begin{equation}
\sum_{l}(A_l^s+A_l^a)=0.
\end{equation}
In the case of a local Fermi Liquid only the $l=0$ parameters
are nonzero and therefore the Landau sum rule reads
\begin{equation}
A_0^s+A_0^a=0
\end{equation}
which leads to Eq.(6) in our paper
\begin{equation}
F_0^s=-F_0^s/(1+2F_0^s).
\end{equation}
We are surprised by their conclusion in 
Ref.~\cite{van Dongen et al.1998} that this last equation must 
be wrong since the first and second order perturbation
approximations\cite{Muller-Hartman1989} disagree with it. Our
result follow necessarily from equations (2), (3), and (4) which
are cornerstones of the theory of metals.  
Their criticism, based on the Hartree-Fock approximation
and second order perturbation
theory implies that if
an exact result contradicts the above approximations then
the exact result must
be wrong.  

Their third comment is concerned with our use of the Pomeranchuk 
criterion for phase separation. The phase separation which is
excluded by our arguments is 
a continuous phase transition from
within the paramagnetic local Fermi Liquid phase.
Of course, we cannot exclude phase separation from 
another phase or through a first-order phase transition.
We also noted in Ref.~\cite{Engelbrecht and Bedell1995} that
any phase transitions at finite momentum, SDW, CDW, etc.
can not be ruled out by our constraints and can not be addressed
within Fermi Liquid Theory.

At the end van Dongen, 
Uhrig, and M\"{u}ller-Hartmann add one more comment on the
$\Phi$-derivability of interacting Fermi systems in 
$d\rightarrow\infty$-dimensions.
Their equation
\[
\frac{\delta^{2}\Phi_{d=\infty}[G(p)]}{\delta G(p')\delta G(p)}\neq
\lim_{d\to\infty}\frac{\delta^{2}\Phi_{d}[G(p)]}{\delta G(p')\delta G(p)}\; ,
\]
implies that the second variational
derivative of the potential $\Phi$ is not a continuous function
of the number of dimensions $d$ which means that 
the limit does not exist. Further
investigation of the subtleties of the many-body theories in
infinite dimensions and their relevance to the properties of
systems in the physical three dimensions may be enlightening. 

We hope that the above explanations will give some peace of mind
to our colleagues and deepen their understanding
of Fermi Liquid Theory and the theory of metals.

We would like to thank G. Kotliar for useful discussions
on these topics. J.R.E. would like to thank
M\"{u}ller-Hartmann and D. Vollhardt for the stimulating discussions
and the Aspen Center for Physics.  
This work was sponsored by Department of Energy grant, DEFG0297ER45636.


\begin{references}

\item[*]
Address after Oct. 1, 1998: Theory of Condensed Matter Group, Cavendish Laboratory,
Madingley Road, Cambridge CB3 0HE, UK.

\bibitem{van Dongen et al.1998}P.G.J. van Dongen$^{(1)}$, G.S. Uhrig and 
E. M\"{u}ller-Hartmann, cond-mat/9807276.

\bibitem{Engelbrecht and Bedell1995} J.R. Engelbrecht and 
K. S. Bedell, Phys.\ Rev.\ Lett.\ {\bf 74}, 4265 (1995).

\bibitem{Blagoev et al.1998a} K.B. Blagoev, J.R. Engelbrecht 
and K. S. Bedell, cond-mat preprint 9709143, to be published in 
Phil.\ Mag.\ Lett.\ (1998).

\bibitem{Migdal1958}  A.B. Migdal, Sov. Phys. JETP {\bf 7}, 996 (1958).

\bibitem{Schrieffer1964}  J.R. Schrieffer, {\it Theory of Superconductivity}
(W.A. Benjamin, Inc, New York, 1964), p154.

\bibitem{Berk and Schrieffer1966}  N.F. Berk and J.R. Schrieffer, Phys. Rev.
Lett. {\bf 17}, 433 (1966).

\bibitem{Berk1966}N.F. Berk, $Dissertation$, 1966.

\bibitem{Blagoev et al.1998b}K.B. Blagoev, J.R. Engelbrecht, and K.S. Bedell,
Phill. Mag. Lett., to be published.

\bibitem{Abrikosov et al.1963}  A.A. Abrikosov, L.P.\ Gorkov, \& I.E.
Dzyaloshinskii, \textit{Methods of Quantum Field Theory in Statistical
Physics}, Dover, 1963.


\bibitem{Landau1957}  L.D. Landau, Zh. Eksp. Teor. Fiz. \textbf{35}, 97
(1958) [Sov. Phys. JETP \textbf{8}, 70 (1959)].

\bibitem{Baym and Pethick1991}  G. Baym and C. Pethick, {\it Landau Fermi-Liquid
Theory}, (John Wilet \& Sons, 1991).

\bibitem{Muller-Hartman1989}E. M\"{u}ller-Hartmann, 
Int. J. Mod. Phys. B {\bf 3}, 2169 (1989).

\end{references}
\end{document}